# Cooperative Lane Changing via Deep Reinforcement Learning

Guan Wang, Jianming Hu, *Member, IEEE*, Zhiheng Li, *Member, IEEE*, Li Li, *Fellow, IEEE*

*Abstract*— In this paper, we study how to learn an appropriate lane changing strategy for autonomous vehicles by using deep reinforcement learning. We show that the reward of the system should consider the overall traffic efficiency instead of the travel efficiency of an individual vehicle. In summary, cooperation leads to a more harmonic and efficient traffic system rather than competition.

## I. Introduction

AUTONOMOUS driving is widely accepted as a major tool to alleviate traffic congestions and reduce traffic accidents [1]. However, it remains challenging for automated vehicles to drive properly with respect to some complex environments.

How to make appropriate lane changes is one of the recent hot topics [2], because bad lane change maneuver is the major cause of many severe highway accidents and traffic jams. Some studies focused on adaptive and efficient lane change trajectory planning for vehicles [3]-[4]. Some studies emphasized the required controller to track the planned trajectory [5]-[6]. Some recent studies collected and examined the naturalistic driving data to learn and mimic human drivers' lane change actions when implementing autonomous driving [7]-[9]. Some latest studies also proposed end-to-end learning technique to directly model the relationship between the video sensing data and the desired lane change actions [10].

In this paper, we study the emerging reinforcement learning (RL) [11]-[12] based lane change. Usually, RL approaches allow an automated vehicle to gradually learn how to drive through interaction with the environment and gain skills from their awards/mistakes. During the last decade, RL had been used to implement many functions of automated vehicles, including lane keeping [13]-[14] and adaptive cruising [15], and researchers have released some RL benchmarks in mixed-autonomy traffic scenarios [16].

Some recent studies had shown interests in RL based lane changing [17]-[20]. Different from end-to-end approaches [10], we are interested to design an RL model to accept neighboring vehicles' position data as input and yield the lane-changing decision as output. More than 10 papers that had been published by both IEEE Intelligent Vehicle Symposium 2018 and 2019 to study related problems. However, most existing studies in this direction consider the reward of the studied vehicle only. For example, DeepTraffic [20] sets the goal to drive the vehicle as fast as possible through dense highway traffic. As pointed out in [21]-[23], inappropriate choice of performance index may urge automated vehicles to compete for their rights of way or take unnecessary over-conservative actions, which therefore leads to unexpected traffic congestions.

To solve this problem, we adopt the concept of cooperative driving and study how to learn an appropriate lane changing strategy for autonomous vehicles by setting proper reward function for reinforcement learning. The idea of cooperative driving is to make well arrangement of vehicle movements so as to make full use of limited road resources and reducing competition [4]-[6]. Especially, we consider both the delay of an individual vehicle and the overall traffic efficiency at the studied road segment in the reward function. Purchasing too short delay often leads to selfish competition lane changing behaviors; while, to reducing the overall delay of vehicles advocates for modest lane changing behavior. Simulation results show that the overall traffic efficiency could be boosted, if these two objectives could be properly mixed.

To better explain our findings, the rest of this paper is arranged as follows. *Section II* presents the deep RL model that is used to implement cooperative lane changing. *Section III* provides the simulator design. *Section IV* gives experiment settings and the simulation results of the proposed model. Finally, *Section V* concludes the paper.

Manuscript received June 16th, 2019. This work was supported in part by the National Natural Science Foundation of China (61790565), Beijing Municipal Science and Technology Commission Program (D171100000317002), Beijing Municipal Commission of Transport Program (ZC179074Z). (Corresponding author: *Li Li*)

G. Wang, J. Hu, and Z. Li are with Department of Automation, Tsinghua University, Beijing, China 100084. Z. Li is also with the Graduate School at Shenzhen, Tsinghua University, Shenzhen 518055, China.
L. Li is with Department of Automation, BNRist, Tsinghua University, Beijing, China 100084. (Tel: +86(10)62782071; Email: li-li@tsinghua.edu.cn)

## II. THE DEEP REINFORCEMENT LEARNING MODEL FOR COOPERATIVE LANE CHANGING

### A. Problem Presentation and the Model

The symbols used in this paper are listed in Table I.

**TABLE I**
**THE NOMENCLATURE LIST**

| Symbol | Definition |
|---|---|
| $M_t^{(i)}$ | traffic snapshot of vehicle $i$ at time $t$ |
| $v_t^{(i)}$ | actual longitudinal speed of vehicle $i$ |
| $v_{y,t}^{(i)}$ | actual lateral speed of vehicle $i$ |
| $x_t^{(i)}$ | longitudinal location of vehicle $i$ |
| $v_{\exp}^{(i)}$ | expected speed of vehicle $i$ |
| $\Delta v_t^{(i)}$ | error between actual and expected longitudinal speed |
| $A$ | decision set of each vehicle |
| $s_t^{(i)}$ | state of vehicle $i$ |
| $r_t^{(i)}$ | reward function of vehicle $i$ |
| $a_t^{(i)}$ | action of vehicle $i$ |
| $q_t$ | traffic flow rate at time $t$ |
| $T_{up}$ | upstream departure interval |
| $\alpha$ | lane changing cooperation coefficient |

In this paper, we formulate the lane-changing of each vehicle as a Markov decision process (MDP) [11]-[12] in terms of a tuple $\langle S, A, R, T, \gamma \rangle$. The tuple is composed of state space $S$, action space $A$, reward function $R$, a transition model $T$ with a transition conditional density $p(s_{t+1} | s_t, a_t)$ satisfying the Markov property, and a discount factor $\gamma$. By maximizing the expected discounted sum of reward $R_t = \sum_{t'=t}^{T} \gamma^{t'} r(s_{t'}, a_{t'})$, the studied vehicle learns a policy $\pi(a_t | s_t)$ to interact with other vehicles on the road via taking the corresponding action $a_t$ based on an observation $s_t$ at time $t$. The action-value function is

$$Q^*(s,a) = \max_{\pi} \mathbb{E}[R_t | s_t = s, a_t = a, \pi] \quad (1)$$

The state space, action space, reward function and deep RL model designs are illustrated respectively as follows.

1) *State Space*:

The state of each vehicle (agent) consists of three sequential frames of traffic snapshots $M_t^{(i)}$ and corresponding speed difference between actual and expected speed. That is

$$s_t^{(i)} = [M_{t-2}^{(i)}, M_{t-1}^{(i)}, M_t^{(i)}, \Delta v_{t-2}^{(i)} \Delta v_{t-1}^{(i)}, \Delta v_t^{(i)}]^T \quad (2)$$

$$\Delta v_t^{(i)} = v_t^{(i)} - v_{\exp}^{(i)} \quad (3)$$

As we assume that there is no support of Vehicle-to-X (V2X) communication and each vehicle only relies on on-board sensors to detect the surrounding traffic situations, complex cooperation mechanism like negotiation [24] might not work well. Therefore, we take the traffic snapshot [20] to help the studied vehicle to learn the surrounding situations.

Inspired by the model proposed in [20], we define traffic snapshot $M_t^{(i)}$ as a two-dimensional occupancy grid that reflects the traffic situation around each vehicle, and $\Delta v_t^{(i)}$ works to indicate the traveling efficiency of vehicle $i$. Fig.1 gives an example of the snapshot of vehicle $i$ (the yellow one). The snapshot is set to "1" for cells where there exists a vehicle and "0" for those empty cells. $s_t^{(i)}$ will be fed into the Deep Q-network as input. The concept of traffic snapshot is suitable for multiple traffic scenarios, instead of depending on road geometry or the number of vehicles around the agent.

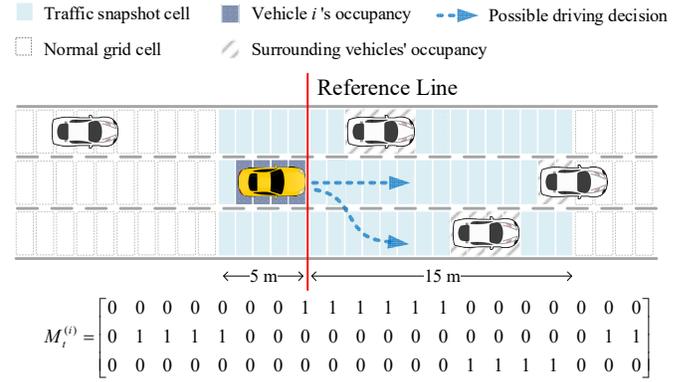

Fig.1. An example of traffic snapshot of vehicle $i$

2) *Action Space*:

Action $a_t^{(i)}$ represents the driving decision of vehicle $i$ and will be checked in the collision detection stage. $a_t^{(i)}$ should be one of the elements of decision set $A$:

$$a_t^{(i)} \in A = \{left, right, speedup, none\} \quad (4)$$

A vehicle can choose to change to left/right lane, speed up with a fixed speed increment $v_{acc} = 0.4$m/s until reaches the maximum speed $v_{\exp}^{(i)}$, or keep the current speed in the current lane. Deceleration is not included in set $A$ because we set a separated collision-check process to modify the speed. The detailed lane-changing model and collision-check processes will be introduced in detail in section III.

3) *Reward Function*:

Reward $r_t^{(i)}$ works as an evaluation of the current state and it is critical to the process of learning the proper action given an observed state in MDP. It determines the degree of cooperation in our lane-changing model.

$r_t^{(i)}$ is a trade-off among vehicle's own traveling efficiency, cooperation with others and the traffic flow rate. As [17]-[19] and [33], we take the widely used linear combination format for the reward function

$$r_t^{(i)} = r_v^{(i)} + r_{cl}^{(i)} + q_t \quad (5)$$

$$r_v^{(i)} = \Delta v_t^{(i)} / (v_{max} - v_{min}) \quad (6)$$

$$r_{cl}^{(i)} = \begin{cases} -\alpha & \text{when changing lane} \\ 0 & \text{otherwise} \end{cases} \quad (7)$$

where $r_v^{(i)}$ is the reward for the studied vehicle's individual traveling efficiency, $r_{cl}^{(i)}$ is the lane changing penalty, $q_t$ is the current traffic flow rate, $\alpha$ is the cooperation coefficient for lane changing behavior, and $v_{\min}$ and $v_{\max}$ are the lower bound and upper bound of speed respectively. Since a vehicle can get best $r_v^{(i)}$ at $v_{\exp}^{(i)}$, it is desirable to travel as fast as possible under $v_{\exp}^{(i)}$. $r_{cl}^{(i)}$ determines the willingness of lane changing behavior of a vehicle and $q_t$ works to emphasize the total traveling efficiency. Noticing that frequent lane changing behavior may decrease the efficiency of the traffic flow, we punish each lane change action by minus $\alpha$ from the reward. A bigger $\alpha$ will lead the vehicle to learn a more modest lane-changing strategy, and unnecessary lane changing can be restrained.

*B. Deep Reinforcement Learning Algorithms*

In the RL models, the agent interacts with the environment through a sequence of observations, actions and rewards, and select actions that could maximize cumulative rewards [28]. However, the high dimension of input and the delay between actions and resulting rewards bring challenges to conventional RL methods [29]. With the help of deep neural networks, deep RL methods show promising potential in solving coupled non-linear control or decision problem [14]-[17].

In this paper, we use Deep Q-network (DQN) [28][29] to learn the effective lane changing decision-making mechanism. The DQN takes $s_t^{(i)}$ as input and outputs the desired driving decision $a_t^{(i)}$. The $Q^*(s,a)$ in Equation (1) is parameterized by a function approximator $Q(s,a;\theta)$ combined with a convolutional neural network (CNN) [30]. To remove correlations in the observation sequence, experience replay is applied by storing the agent's experience $e_t = (s_t, a_t, r_t, s_{t+1})$ in a dataset $D_t = \{e_1,...,e_t\}$ in DQN [28].

When learning the model, samples are drawn uniformly from $D_t$ to calculate the following loss function (TD error), and parameters are updated using stochastic gradient descent:

$$Loss = \mathbb{E}\left[\left(r_t + \gamma \max_{a_{t+1}} \hat{Q}(s_{t+1}, a_{t+1}) - Q(s_t, a_t)\right)^2\right] \quad (8)$$

where $Q$ denotes the parameters of online Q-network and $\hat{Q}$ denotes the target Q-network. $\hat{Q}$ is updated by $Q$ every $N$ steps and remains the same in other intervals. When making lane-changing decisions, the agent will select and execute an action according to an $\epsilon$ – greedy policy based on the Q-value from the output of the DQN.

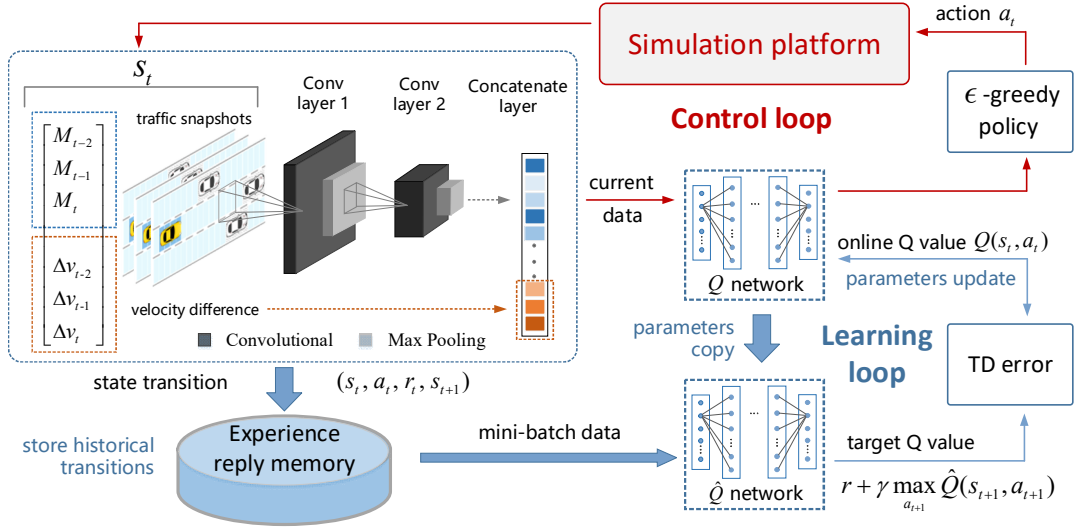

Fig.2. Cooperative lane-changing DQN architecture

**TABLE II**
LAYOUT OF LANE-CHANGING DECISION DQN

| Layer | Parameters |
|---|---|
| Convolutional Layer 1 | patch size = (2,2), stride = (1,2) number of filter = 16 activation = ReLU |
| Convolutional Layer 2 | patch size = (2,2), stride = (1,2) number of filter = 32 activation = ReLU |
| Concatenating Layer | |
| Fully Connected Layer 1 | number of units = 500 activation = ReLU |
| Fully Connected Layer 2 | number of units = 100 activation = ReLU |

The detailed layout of our cooperative lane-changing decision DQN is listed in TABLE II, and the whole architecture is shown in Fig.2. In our model, sequential traffic snapshots from $s_t^{(i)}$ are processed by a 2-layer CNN first, and then concatenated with sequential $\Delta v^{(i)}$ by the concatenating layer. Then the data is fed into the 2-layer fully connected Q-network. Finally, the DQN will output an action $a_t^{(i)}$ as the driving decision proposal.

Since the sequential traffic snapshots can be treated as 3-frame 3x20-binary-pixel graphs, the simple 2-layer CNN could work well for feature abstraction. Some other deep RL structures (e.g. DDPG [38]) with deeper architecture have been tested, and results indicate that adding more layers does not contribute much to the experiment results, considering the increase in calculation time. In addition, as the action space is mainly composed of low dimensional discrete decision choices and the state space is refined before the Q-network, DQN could work well for our lane-changing model. Therefore, the proposed network architecture is chosen to handle the lane-changing decision problem.

## III. SIMULATION PLATFORM

In order to verify the proposed cooperative lane-changing model, we establish a simulation platform. The pipeline of the platform can be summarized as the following steps:

1) Generate new vehicles at the beginning of the road based on an upstream inflow rate.

2) Obtain the environmental data and get driving decision from the proposed lane-changing model.

3) Calculate each vehicle's proper speed, and execute the driving decision.

4) Execute collision-check process and update the locations of all vehicles.

In step 3), the longitudinal and lateral speed will be calculated by the car-following model and lane-changing model respectively in each iteration. A collision-check process will be executed afterward in step 4) to modify longitudinal speed for safety.

### A. Car-Following Model

The car-following model sets a longitudinal speed limit to each vehicle [26][27]. Since the output $a_t^{(i)}$ of the proposed lane-changing model is the desired action value, the actual speed should be modified for safety before applied to update locations. We adopt the simple Newell Car Following model [35] to limit the longitudinal speed of each vehicle for safety:

$$v_{t+1}^{(i)} = \min\left\{v_t^{(i)} + a_t^{(i)}, v_{exp}^{(i)}\left(1 - \exp(-\frac{c\Delta x_t^{(i)}}{v_{exp}^{(i)}} - d)\right)\right\} \quad (9)$$

where $\Delta x_t^{(i)}$ is the longitudinal distance between vehicle $i$ and the longitudinally nearest vehicle around it in the same lane. It should be noted that $a_t^{(i)}$ is either $v_{acc}$ or 0 in the longitudinal direction.

### B. Lane-Changing Model

The lane-changing model gives the lateral speed for the lane-changing vehicles:

$$v_{y,t}^{(i)} = W_{lane} / t_{change} \quad (10)$$

where $W_{lane}$ is the width of the lane and $t_{change}$ is the time for lane changing. To simplify the model, $t_{change}$ is set to 4 seconds uniformly [36]. As the lateral trajectory is not the main factor in our simulation, the lateral speed is set as constant while a vehicle changing lane. We assume that once a lane changing occurs on the studied vehicle $i$, no other lane changing actions are permitted until the time interval elapses [34].

### C. Collision-Check Process

To avoid occasional accidents that may be caused by limited deceleration rate generated by the Newell car-following model, we add an additional collision check to adjust the longitudinal speed of each vehicle as

$$\text{if } \Delta x_t^{(i)} \leq a_1 L_{car} + a_2 v_t^{(i)} \text{ then } v_{t+1}^{(i)} \leq v_t^{(pre)} - v_{cc} \quad (11)$$

where $v_{cc}$ denotes the additional gain for safety, $a_1$, $a_2$ are constant coefficients, and $v_t^{(pre)}$ is the longitudinal speed of the nearest preceding vehicle for the studied vehicle.

## IV. TESTING RESULTS

To illustrate the efficiency of the proposed RL cooperative lane-changing model, we design two simulation scenarios: traffic accident bottleneck (scenario I) and non-accident scenario (scenario II). Fig.3 shows the three-lane highway for simulation experiments. The red dashed line represents the traffic accident bottleneck, which is set referring to [32]. The upstream traffic is set as free flow and vehicles are generated according to the upstream departure interval $T_{up}$.

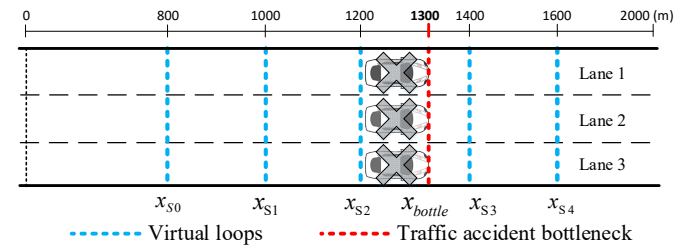

Fig.3. Placement of virtual loops and traffic accident bottleneck

Traffic flow characteristics are detected by virtual loops in the simulator as [31]. The placement of virtual loops is shown in Fig.3 as a blue dashed line, and the loops are placed with an equal interval of 200 m.

Similar to [20], we only consider highway scenarios which are discretized into a grid with a scale of 1 m in width. A line in front of vehicle $i$ is set as the reference, and the traffic environment around vehicle $i$ (15m ahead and 5m behind the reference line) is taken into account for traffic snapshot. The length of each vehicle is set to $L_{car} = 4$ m uniformly in these scenarios. We set $a_1 = 2$, $a_2 = 0.5$ and $v_{cc} = 1$ m/s for Equation (11). Each vehicle is generated with a random expected speed $v_{\exp}^{(i)}$ to achieve.

The RL based cooperative lane changing model is trained in scenario II for 90,000 steps and then tested in scenarios I. As for other hyper-parameters of our DQN, the learning rate is 0.01, the reward discounted factor $\gamma$ is 0.9, the memory size for experience replay is 2000 samples, and $\epsilon$ – greedy threshold is 0.9. The target Q-network updates every $N = 500$ steps with the parameters from the online Q-network.

### A. Traffic Accident Scenario (Scenario I)

This scenario works to show how the cooperation mechanism improves the overall traffic efficiency when an accidental traffic bottleneck occurs, where the moving jam may propagate upstream and result in a reduction of throughput [32]. In this scenario, the traffic accident bottleneck will last about 50,000 steps (1.25 hours) at $x_{bottle} = 1300$ m according to Fig.3, and the whole scenario will last 90,000 steps (2.5 hours). During the period, vehicles around $x_{bottle}$ in all lanes are set to be stuck with a ratio $p$ and other vehicles will be released. $p$ will gradually decay from 1 to 0. Here, we take $\alpha = 8$ in Equation (7) as a more cooperative lane-changing strategy, and $\alpha = 0$ as a non-cooperative strategy. Approximate choice of $\alpha$ will be explored in scenario II.

To give an intuitive illustration of the lane-changing results of the proposed model, we choose partial vehicle trajectories on lane 1 from the simulation results, as shown in Fig.4. The obvious stop-and-go waves in Fig.4 (a) cause traffic jams, and travel backward along the road and reduce road capacity. On the contrary, in Fig.4 (b) the stop-and-go waves are eliminated effectively by a cooperative lane-changing mechanism, and road resources get better usage by filling white spaces [25]. Since lane changing could trigger stop-and-go waves [37], the limit of unnecessary lane changing would improve traffic flow. The traffic bottleneck is controlled more effectively with the cooperation mechanism, since the speed of vehicles in Fig.4 (b) reaches above 15 m/s after 74,000 simulation steps while the bottleneck remains in Fig.4 (a).

Fig.5 shows the resulting flow-density relations of the models trained respectively with the non-cooperative and the more cooperative strategy. It should be noted that the models are trained in the non-accident scenario and directly applied to this accident bottleneck scenario without modification. The superior performance of cooperation over non-cooperation might imply the transferability of the proposed model.

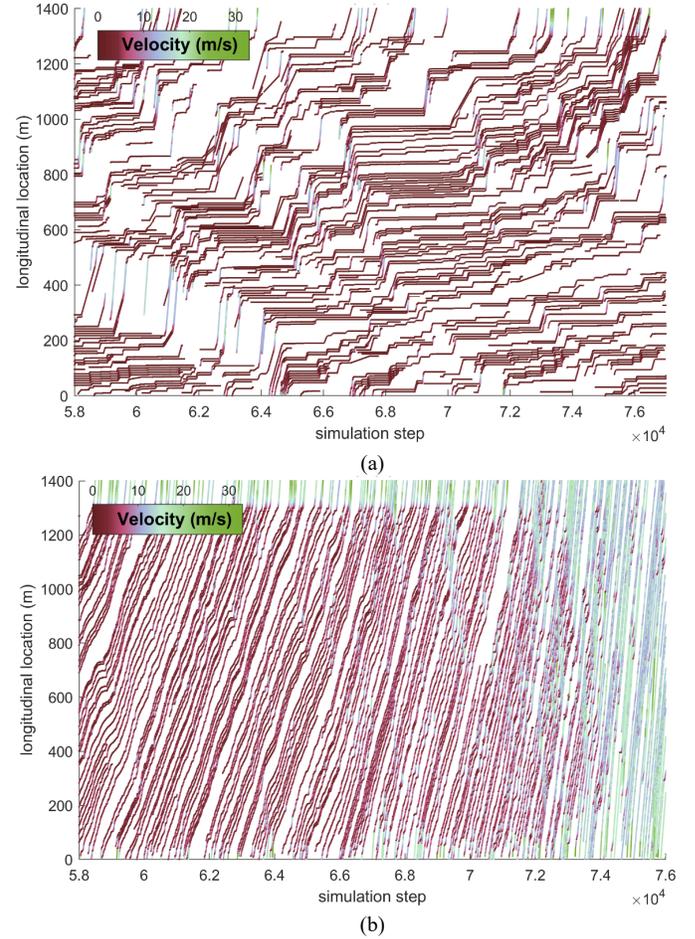

Fig.4. Partial vehicle trajectories on lane 1 with different strategy settings: (a) sampled from the model trained with the non-cooperative strategy, (b) sampled from the model trained with a more cooperative strategy

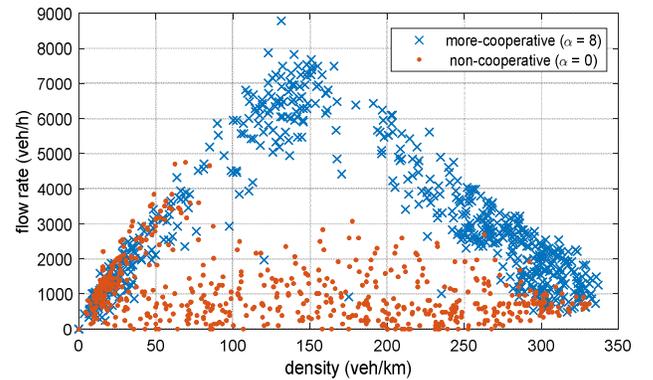

Fig.5. Flow-density relations detected by the virtual loops under different lane changing strategies

### B. Non-Accident (Scenario II)

In Scenario II, we will discuss how to choose a proper cooperation coefficient $\alpha$ in different traffic conditions. This scenario is set similar to Fig.3, but the accident bottleneck is removed to help us focus on the influence of $\alpha$ in free flow.

Since the inflow rates and lane number could have a significant influence on the simulation result, we will focus on multiple lane numbers and inflow rates conditions. We examine a set of $\alpha$ to find the optimal $\alpha$ in each condition:

$$\alpha = \{0, \frac{1}{8}, \frac{1}{4}, \frac{1}{2}, 1, 2, 4, 8, 16, 24, 32, 48\} \qquad (12)$$

We take the mean traveling speed as the criteria for choosing the optimal $\alpha$, denoted as $\alpha^*$. Each condition is repeated for 10 times, and the results are then averaged to reduce the impact of randomness. Since the model focuses on the vehicles on the highway, all $v_{exp}$ are clipped within 40 km/h to 110 km/h, according to the statistical data in [39].

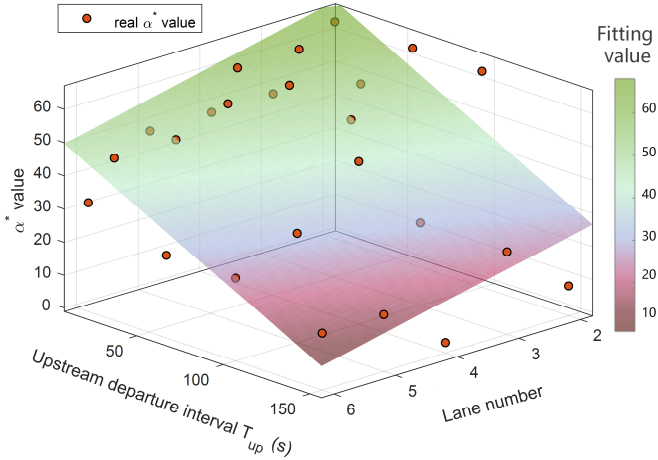

Fig.6. $\alpha^*$ in different inflow intervals and lane numbers and the resulting fitting hyperplane, where the red points denote real $\alpha^*$ values.

Fig.6 shows the $\alpha^*$ under different lane numbers (from 2 lanes to 6 lanes) and inflow rates, and a fitting hyperplane of the $\alpha^*$ is shown to give an intuitive demonstration about the trend of $\alpha^*$. According to Fig.6, optimal $\alpha^*$ will increase as the departure interval decreases and lane number decreases. In other words, cooperation and modesty should be valued more if there are not many lanes in the road or the inflow rates are high, and a bigger $\alpha^*$ would perform better in these conditions.

Fig.6 also indicates that $\alpha$ is a subtle trade-off between individual and overall traveling efficiency. It is hard to find a simple rule of $\alpha$ selection in all conditions, because of the dynamicity of traffic flow and the complex interaction between flow rate and lane change.

## V. CONCLUSIONS

In this paper, we propose a lane-changing model based on deep RL method. Instead of only focusing on the interest of an individual vehicle, the proposed model emphasizes overall traffic efficiency by introducing a cooperation mechanism into the reward function. Testing results show that the cooperation mechanism could improve the overall traffic efficiency especially in congested conditions, and lead to a more harmonic traffic system.

Further research may involve two interesting directions. First, we would focus on how to learn an adaptive mechanism for choosing cooperation coefficient $\alpha$ in different traffic conditions or calculate $\alpha$ more efficiently. Besides, we are also interested to explore how to establish a more efficient representation of vehicles' state based on V2X assumptions, which might accelerate the searching and training of the DRL models.